\documentclass[11pt,a4paper]{article}
\usepackage{graphicx}
\usepackage{amssymb}
\usepackage{amsmath}
\usepackage{amsthm}
\usepackage{amsbsy}
\usepackage{cite}
\usepackage{verbatim} 
\usepackage{capt-of}
\usepackage{arydshln}
\usepackage[dvipsnames]{xcolor}

\newcommand{\bEq}{\begin{equation}}
\newcommand{\eEq}{\end{equation}}
\newcommand{\bEQ}[1]{\begin{equation} \begin{array}{#1}}
\newcommand{\eEQ}{\end{array} \end{equation}}

\newcounter{CorrectThis}

\begin{document}
	
\pagestyle{empty}	
\null	
\vfill
	
\begin{center}
			
{\Large  {\bf Grain boundary energy models and boundary splitting} }\\

\vskip 1.0cm
A. Morawiec
\vskip 0.2cm 
{Institute of Metallurgy and Materials Science, 
Polish Academy of Sciences, \\ 
Krak{\'o}w, Poland.
}
\\
E-mail: nmmorawi@cyf-kr.edu.pl \\
Tel.: ++48--122952854, \ \ \  Fax: ++48--122952804 \\			
\end{center}

\vfill

\noindent
{\bf Abstract}
\\
Models of grain boundary energy are essential for 
predicting the behavior of polycrystalline materials. 	
Typical models 
represent the minimum boundary energy as a function 
of macroscopic boundary parameters.
An energy model may allow for boundary dissociation, 
i.e., for a further reduction of the overall energy by splitting 
a boundary into two boundaries parallel to the original one.
Such splitting is prevented by constraining the 
energy model with inequalities opposite to the boundary wetting condition.
The inequalities are applicable only to triplets of boundaries 
that match the assumed geometric configuration.
Relationships connecting the parameters of such boundaries are derived, 
implications of the inequalities 
that prevent boundary splitting are considered,
and an example energy model is shown to allow boundary decomposition. 
Knowing whether a given energy model permits boundary 
dissociation and which boundaries can be affected 
is important for evaluating its performance in polycrystal simulations.

	\vskip 0.5cm
	

	
\newpage	
\pagestyle{plain}

\noindent
There is a growing awareness 
of the need to develop advanced grain boundary 
energy models that may be useful in simulations
of polycrystalline materials and ultimately in 
explaining their behavior.
Typical models (e.g., 
\cite{bulatov2014grain, sarochawikasit2021grain,
chirayutthanasak2024universal, Morawiec_2025a, Ratanaphan_2019, zhang2022molecular})
are based on molecular statics simulations at 0K 
(e.g.,  
\cite{olmsted2009survey,Ratanaphan_2015,Ratanaphan_2019,zhang2022molecular})
and the represent minimum energy 
as a single-valued function of macroscopic boundary parameters
\cite{Goux_1961,Lange_1967,Fortes_1972}.\footnote{The analysis 
presented below can also be applied to experimentally obtained free energy functions 
(e.g.,  \cite{saylor2003relative,beladi2014five,shen2019determining}),
but it will be limited by the low resolution of the experimental data.}	
When seeking minimum energy configurations, 
one may look beyond the 
simplest solutions.
In principle, with arbitrary boundary energies,
a structure can lower its overall energy by 
replacing a high-energy boundary with several 
boundaries whose combined energy is less than 
that of the original boundary; see, e.g., 
\cite{Goodhew_1978,Forwood_1984,Sutton_1995}.
Exploring every possible configuration would be difficult. 
However, one can consider 
the specific case 
-- consistent with the macroscopic 
description -- 
in which a planar boundary 
is replaced by two planar boundaries parallel to the original one.
Given a grain boundary energy model, 
the question is whether it allows such replacements. 
The formal condition prohibiting such events is 
the opposite of the boundary wetting condition.
It applies to boundaries in the assumed 
geometric configuration, i.e., the macroscopic parameters of these 
boundaries are interrelated due to geometric constraints.

This note discusses implications for energy modeling 
arising from the assumption that a lower-energy 
state cannot be reached by splitting a grain boundary 
into two parallel boundaries.
To determine whether an energy model meets this assumption, 
the explicit form of the geometric compatibility conditions is needed.
These are first derived and then used to test 
the celebrated Bulatov–Reed–Kumar (BRK) 
energy model \cite{bulatov2014grain}.
The susceptibility of a boundary to splitting
is quantified by a dimensionless parameter.
Before addressing the main topic, preliminary information on definitions, conventions, and notation is provided.

Crystal orientations and misorientations are identified with rotations 
relating Cartesian reference frames ascribed to the crystals and the sample. 
A rotation about $\mathbf{k} = [k^1\ k^2\ k^3]$ by angle $\theta$ is represented by the special orthogonal matrix 
$R(\mathbf{k},\theta) = 
I  \cos \theta +\mathbf{k} \otimes \mathbf{k} \, (1- \cos \theta) + 
\text{Skew}(\mathbf{k}) \, \sin \theta$, where $I$ is the identity matrix,  
the $ij$-th entry of the matrix $\text{Skew}(\mathbf{k})$ is  
$\sum_l \varepsilon_{ijl} k^l$, and $\varepsilon$ is the permutation symbol.
With orientations of crystals $1$ and $2$ represented by  
matrices $g_1$ and $g_2$, respectively, the misorientation between these 
crystals is represented by $M=g_1 g_2^{-1}$.  
The boundary plane is characterized by vector $\mathbf{n}$, 
normal to the plane, directed outward from the first crystal, 
and expressed in the coordinate system of that crystal. 
The boundary is macroscopically defined by the pair $(M, \mathbf{n})$.
Equivalently, it can be described by an interface 
matrix $\mathbf{B}$ constructed from $M$ and $\mathbf{n}$;
see \cite{Morawiec_1998,Morawiec_2009} for details. 
With  $(M, \mathbf{n}) \simeq \mathbf{B}$ representing 
a boundary between grains $1$ and $2$, 
the boundary between grains $2$ and $1$ is represented by
$(M^T, -M^T \mathbf{n})  \simeq  \mathbf{B}^T$.
Due to crystal symmetry, the $\mathbf{B}$ matrix 
corresponds to the same physical boundary as 
$\mathbf{B}^-  \simeq  (M, -\mathbf{n})$
and 
$\mathbf{C}_1 \mathbf{B} \mathbf{C}_2^T$, where the special orthogonal matrices 
$\mathbf{C}_1$ and $\mathbf{C}_2$ represent proper rotations
of the crystal point group \cite{Morawiec_1998,Morawiec_2009}.
The parameters $(I,\mathbf{n}) \simeq \mathbf{B}_0$ correspond to 
the unique 'no-boundary' case.
Parameters of boundary types used below 
are listed in Table~\ref{table:1}.
Boundary energy $\Gamma$ is assumed to be a 
continuous function of macroscopic parameters.
The function takes the same values for 
symmetry-equivalent boundary representations, i.e., 
$
\Gamma(\mathbf{B}) = \Gamma(\mathbf{C}_1 \mathbf{B} \mathbf{C}_2^T)= \Gamma(\mathbf{B}^-) = \Gamma(\mathbf{B}^T)  
$ \cite{Morawiec_1998},
and is positive except at 
points equivalent to $\mathbf{B}_0$, 
where it equals zero.

\begin{table}[t]
	\centering
	\begin{tabular}{l|cll|ll}
 &	$\mathbf{k}$ & $\theta$ & $\mathbf{n}$ 	& $-M^T \mathbf{n}$	&  \\ 
	\hline
$\mathbf{B}_0$ \ &   & $0$ &   &   & $\Sigma 1$ 'no-boundary' \\  		
$\mathbf{B}_1$	& $[1 \, 1\, 1]$ & $\pi/3$ & $[1 \, 1 \, 1] $ \  & 
  $[\overline{1}\, \overline{1}\, \overline{1}]$ & $\Sigma 3$ twin boundary\\  		
$\mathbf{B}_2$	& $[1\, 1\, 0]$ & $\arccos(1/3)$ & $[1 \, 1 \, 1] $ 
	& $[\overline{5} \, \overline{1} \, \overline{1}]/3$ & $\Sigma 3$ \\  		
$\mathbf{B}_3$	& $[3\, 2\, 1]$ & $\arccos(-5/9)$ & $[1 \, 1 \, 1] $ 
	& $[\overline{5} \, \overline{1} \, \overline{1}]/3$ & $\Sigma 9$ \\  		
$\mathbf{B}_4$	& $[1\, 1\, 0]$ & $\arccos(7/11)$ & $[\overline{1} \, 1 \, 3]$ 
	& $[\overline{1} \, 1 \, \overline{3}]$ & $\Sigma 11$ \\  		
$\mathbf{B}_5$	& $[1\, 1\, 1]$ & $\arccos(13/14)$ & $[1 \, 1 \, 1]$ 
	& $[\overline{1} \, \overline{1} \, \overline{1}]$ & $\Sigma 21a$ \\  		
$\mathbf{B}_6$	& $[1\, 1\, 1]$ & $\arccos(59/62)$ & $[1 \, 1 \, 1]$ 
	& $[\overline{1} \, \overline{1} \, \overline{1}]$ & $\Sigma 31a$ \\  		 
	\end{tabular}
	\caption{Boundary types referred to in the text.
Each is represented by the pair $(M, \mathbf{n})$, 
where $M=R \left(\mathbf{k}, \theta \right)$.  
The crystal point symmetry is assumed to be $m\overline{3}m$.
	}
	\label{table:1}
\end{table}

\begin{figure}[t]
	\begin{picture}(300,135)(0,0)
		\put(135,10){\resizebox{7.0 cm}{!}{\includegraphics{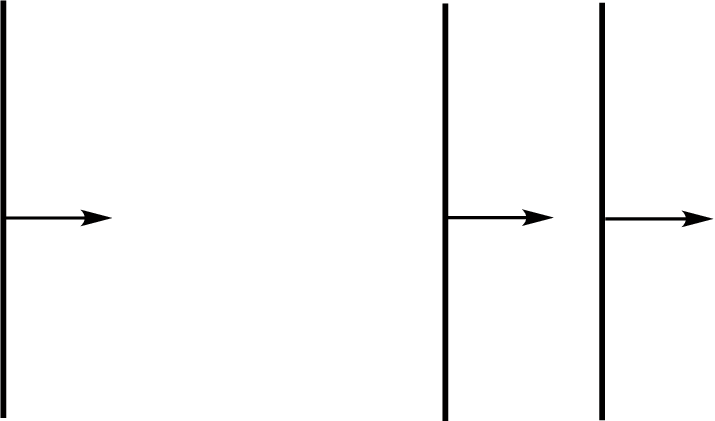}}}
		\put(131,0){$\mathbf{B}_a$ }
		\put(253,0){$\mathbf{B}_{b_1}$ }
		\put(299,0){$\mathbf{B}_{b_2}$ }
		\put(145,71){$\mathbf{n}_{a}$ }
		\put(266,71){$\mathbf{n}_{b_1}$ }
		\put(310,71){$\mathbf{n}_{b_2}$ }
		\put(114,30){$g_1$ }
		\put(146,30){$g_2$ }
		\put(238,30){$g_1$ }
		\put(314,30){$g_2$ }
		\put(276,30){$g_0$ }
		\put(113,130){$M_a=g_1 g_2^{-1}$ }
		\put(225,130){$M_{b_1}=g_1 g_0^{-1}$ }
		\put(296,130){$M_{b_2}=g_0 g_2^{-1}$ }
		\put(97,100){grain 1}
		\put(142,100){grain 2}
		\put(222,100){grain 1}
		\put(311,100){grain 2}
	\end{picture}
	\vskip 0.0cm
	\caption{
		Schematic  illustration
		of decomposition of boundary  
		$\mathbf{B}_a \simeq (M_a,\mathbf{n}_a)$ into the pair of boundaries
		$(\mathbf{B}_{b_1},\mathbf{B}_{b_2})$, where 
		$\mathbf{B}_{b_i} \simeq (M_{b_i},\mathbf{n}_{b_i})$. 
		The symbols $g_k$ ($k=0,1,2$) denote grain orientations.
	}
	\label{Fig_bd_decomp}
\end{figure}
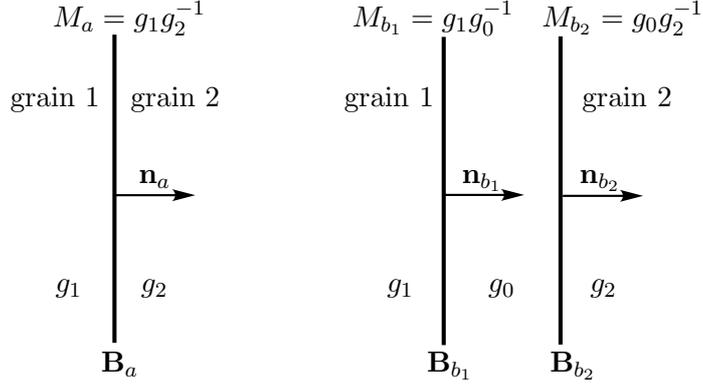

Typical simulations aimed at energy determination
do not allow for formation of new grains 
at the boundary between two originally existing grains.
However, if the energy function is limited only by 
the constraints mentioned above,
formation of such grains is theoretically possible.
The formal condition preventing reaching a lower energy state 
by dissociating 
the boundary $\mathbf{B}_a$ into 
two boundaries $\mathbf{B}_{b_i}$ ($i=1,2$) 
has the form 
\bEq
\Gamma(\mathbf{B}_a) < \Gamma(\mathbf{B}_{b_1})+\Gamma(\mathbf{B}_{b_2}) \  ,
\label{eq:dewetting_condition}
\eEq 
i.e., it is opposite to the boundary wetting condition; see, e.g., 
\cite{Sutton_1995}.
Wettability is quantitatively described be the dimensionless parameter   
$$
w=\max \left\{ 0, \  
1-(\Gamma(\mathbf{B}_{b_1})+\Gamma(\mathbf{B}_{b_2}))/\Gamma(\mathbf{B}_a) 
\right\} \ . 
$$
For the 
boundary $\mathbf{B}_a \simeq (M_a,\mathbf{n}_a)$
to be replaceable by the pair 
$\left( \mathbf{B}_{b_1}, \mathbf{B}_{b_2} \right)$, 
where 
$\mathbf{B}_{b_i} \simeq (M_{b_i},\mathbf{n}_{b_i})$ 
and the planes of $\mathbf{B}_{b_i}$
are parallel to the plane of $\mathbf{B}_a$, 
the parameters of the boundaries must satisfy geometric 
compatibility conditions.
Let the grains forming $\mathbf{B}_a$ have the 
orientations $g_1$ and $g_2$, and 
let the orientation of the grain between 
the boundaries $\mathbf{B}_{b_1}$ and $\mathbf{B}_{b_2}$ be $g_0$
(Fig.~\ref{Fig_bd_decomp}).
One has 
$M_a = g_1 g_2^{-1}$, $M_{b_1} = g_1 g_0^{-1}$ and $M_{b_2} = g_0 g_2^{-1}$.
Hence, the misorientations $M_a$ and $M_{b_i}$
are related by 
\begin{subequations}\label{eq:compatibility_conditions}
	\begin{equation}
		M_a = M_{b_1} M_{b_2} \ .
	\end{equation}
	The vectors $\mathbf{n}_a$ and $\mathbf{n}_{b_1}$ are given in the same reference frame (of crystal 1) and have the same components, i.e., 
	\begin{equation}
		\mathbf{n}_{b_1}=\mathbf{n}_a \ .
	\end{equation}
	Similarly, the vector $-M_{b_2}^T \mathbf{n}_{b_2}$ 
	is equal to $-M_a^T \mathbf{n}_a$, 
	and hence, 
	\begin{equation}
		\mathbf{n}_{b_2} = 		
		M_{b_1}^T \mathbf{n}_a \ .
	\end{equation}
\end{subequations}
It is easy to verify that if $\mathbf{B}_a$ is replaceable by  
$\left( \mathbf{B}_{b_1}, \mathbf{B}_{b_2} \right)$
in the sense that $\mathbf{B}_a$ and $\mathbf{B}_{b_i}$ satisfy 
relations (\ref{eq:compatibility_conditions}),
then
$\mathbf{B}_a^-$ is replaceable by  
$\left( \mathbf{B}_{b_1}^-, \mathbf{B}_{b_2}^- \right)$, 
$\mathbf{B}_a^T$ is replaceable by  
$\left( \mathbf{B}_{b_2}^T, \mathbf{B}_{b_1}^T \right)$
and 
$\mathbf{C}_1 \mathbf{B}_a \mathbf{C}_2^T$ 
is replaceable by 
$\left( \mathbf{C}_1 \mathbf{B}_{b_1}  \mathbf{C}_0^T, \
\mathbf{C}_0 \mathbf{B}_{b_2} \mathbf{C}_2^T \right)$.
This means that relations (\ref{eq:compatibility_conditions}) 
are consistent with the symmetry requirements.
The configuration with 
$\mathbf{B}_a$ replaceable by the pair 
$\left( \mathbf{B}_{b_1}, \mathbf{B}_{b_2} \right)$
is fully determined by complete data for one boundary and the misorientation 
of one of the other two boundaries,
e.g., by 
$\mathbf{B}_a$ and $M_{b_2}$.
Accordingly, the wettability parameter $w$ can be seen as a function of, 
say, $\mathbf{B}_a$ and $M_{b_2}$.
The quantity $w(\mathbf{B}_a,M_{b_2})$ is a measure of how susceptible 
$\mathbf{B}_a$ is to being replaced by two parallel boundaries, 
the second of which has the misorientation $M_{b_2}$
(and $\mathbf{n}_{b_2}$ and $\mathbf{B}_{b_1}$ can be determined from 
(\ref{eq:compatibility_conditions})).

A comprehensive verification of whether a given energy function meets
the condition 
$w(\mathbf{B}_a,M_{b_2})=0$ for every pair $(\mathbf{B}_a,M_{b_2})$
would be computationally expensive. 
However, it is easy to see that 
with typical character of the energy function, 
this condition can be violated 
only in regions where at least one of $\Gamma(\mathbf{B}_{b_i})$ is low,  
and it is enough to check these cases.

Inequality (\ref{eq:dewetting_condition}) certainly 
matters when $\mathbf{B}_{b_1}$, $\mathbf{B}_{b_2}$, or both are 
small-angle boundaries, 
and some of its implications are easy to deduce.
Let
$\mathbf{B}_a$ and 
$\mathbf{B}_{b_i}$ ($i=1,2$)
be small-angle twist boundaries with the same twist axis and twist direction,  
i.e., 
$M_{b_i} = R(\mathbf{k},\theta_{b_i})$,
$M_a = R(\mathbf{k},\theta_a)$, $\mathbf{n}_a=\mathbf{k}=\mathbf{n}_{b_i}$,
$\theta_{b_i}>0$
and  $\theta_a = \theta_{b_1}+\theta_{b_2}$.
These parameters satisfy the compatibility conditions  
(\ref{eq:compatibility_conditions}).
Assuming $\mathbf{k}$ is fixed, only the misorientation angles are variable,  
and (with some abuse of notation) 
inequality (\ref{eq:dewetting_condition}) can be written in terms 
of the misorientation angles as
$\Gamma(\theta_a) < \Gamma(\theta_{b_1})+\Gamma(\theta_{b_2})$.
This means that in the case of twist boundaries with fixed misorientation axis,
if inequality (\ref{eq:dewetting_condition}) holds, then
$\Gamma$ as a function of the misorientation angle $\theta$
is subadditive 
near $\theta=0$.\footnote{The class of subadditive functions is related to, and includes, the class of concave functions \cite{Bruckner_1962}. 
Notably, the Read–Shockley function is strictly concave and therefore also subadditive.}


The above scheme can be generalized.
Condition (\ref{eq:dewetting_condition})  
puts a limit on the slope of energy change
in a neighborhood of an arbitrary boundary  
$\mathbf{B}_C \simeq (M_C,\mathbf{n}_C) \simeq \mathbf{B}_{b_1}$.
Let $\mathbf{B}_{\epsilon} \simeq (M_{\epsilon},\mathbf{n}_{\epsilon}) \simeq  \mathbf{B}_{b_2}$ 
be a small angle boundary with  
$\mathbf{n}_{\epsilon} = M_C^T \mathbf{n}_C$.
Based on (\ref{eq:compatibility_conditions}),
inequality (\ref{eq:dewetting_condition}) constrains energy of
at $\mathbf{B} \simeq  
( M_C M_{\epsilon}, \mathbf{n}_C) \simeq \mathbf{B}_a$. 
With small-angle $M_{\epsilon}$, $\mathbf{B}$ is close 
to $\mathbf{B}_C$
(as measured by boundary distance \cite{Morawiec_2019_AM}).
Using the new symbols, the general expression (\ref{eq:dewetting_condition}) 
is restated as 
\bEq
\Gamma(\mathbf{B}) < \Gamma(\mathbf{B}_C) + \Gamma(\mathbf{B}_{\epsilon}) \ , 
\label{eq:dewet_cond_small_angle}
\eEq
so it can be referred to in special cases when the term 
$\Gamma(\mathbf{B}_{\epsilon})$ is a low energy of a low-angle boundary. 
With a fixed center $\mathbf{B}_C$ and varying $M_{\epsilon}$, 
condition (\ref{eq:dewet_cond_small_angle}) pertains directly 
to boundaries $\mathbf{B}$  characterized by the fixed plane normal $\mathbf{n}_C$
and the family of misorientations of the form $M_C M_{\epsilon}$.
If there is
a cusp at $\mathbf{B}_C$, 
inequality (\ref{eq:dewet_cond_small_angle}) implies that 
cusp slopes cannot be steeper than corresponding slopes of 
the cusp at $\mathbf{B}_0$.

Grain boundary energy models are tacitly assumed to satisfy (\ref{eq:dewetting_condition})
for all $\mathbf{B}_a$ replaceable 
by $\left( \mathbf{B}_{b_1}, \mathbf{B}_{b_2} \right)$.
However, there seems to be no reason to believe that any of the 
existing models have been tested in this respect.
For illustration, the BRK model \cite{bulatov2014grain} is considered.
According to this model for Ni, 
the energies of the boundaries 
\bEq
\mathbf{B}_a  = \mathbf{B}_3  \ , \ \ \ \  
\mathbf{B}_{b_1}  = \mathbf{B}_1  \ \  \mbox{and} \ \  
\mathbf{B}_{b_2}  = \mathbf{B}_2  \ 
\label{eq_boundaries_S9}
\eEq  
are $1.040 \mbox{J}/\mbox{m}^2$, 
$0.062 \mbox{J}/\mbox{m}^2$
and $0.874 \mbox{J}/\mbox{m}^2$, respectively.
Since 
$
\Gamma(\mathbf{B}_{b_1}) + \Gamma(\mathbf{B}_{b_2})  \approx 
0.936 \mbox{J}/\mbox{m}^2 < 1.040 \mbox{J}/\mbox{m}^2 \approx 
\Gamma(\mathbf{B}_a) 
$, inequality (\ref{eq:dewetting_condition}) is violated 
and $w \approx 0.10 > 0$.
Thus, the model permits  the $\Sigma 9$ boundary with the planes 
$(1 \, 1 \, 1)  ||  (\overline{5} \, \overline{1} \, \overline{1})$
to split into two parallel boundaries: the twin boundary and 
the $\Sigma 3$ boundary with the planes
$(1 \, 1 \, 1)  ||  (\overline{5} \, \overline{1} \, \overline{1})$.
Due to the continuity of the energy function,
condition (\ref{eq:dewetting_condition}) is violated 
not only at $\mathbf{B}_a  = \mathbf{B}_3$ but also 
in its neighborhood.

\begin{figure}
	\begin{picture}(300,185)(0,0)
		\put(0,2){\resizebox{7.0 cm}{!}{\includegraphics{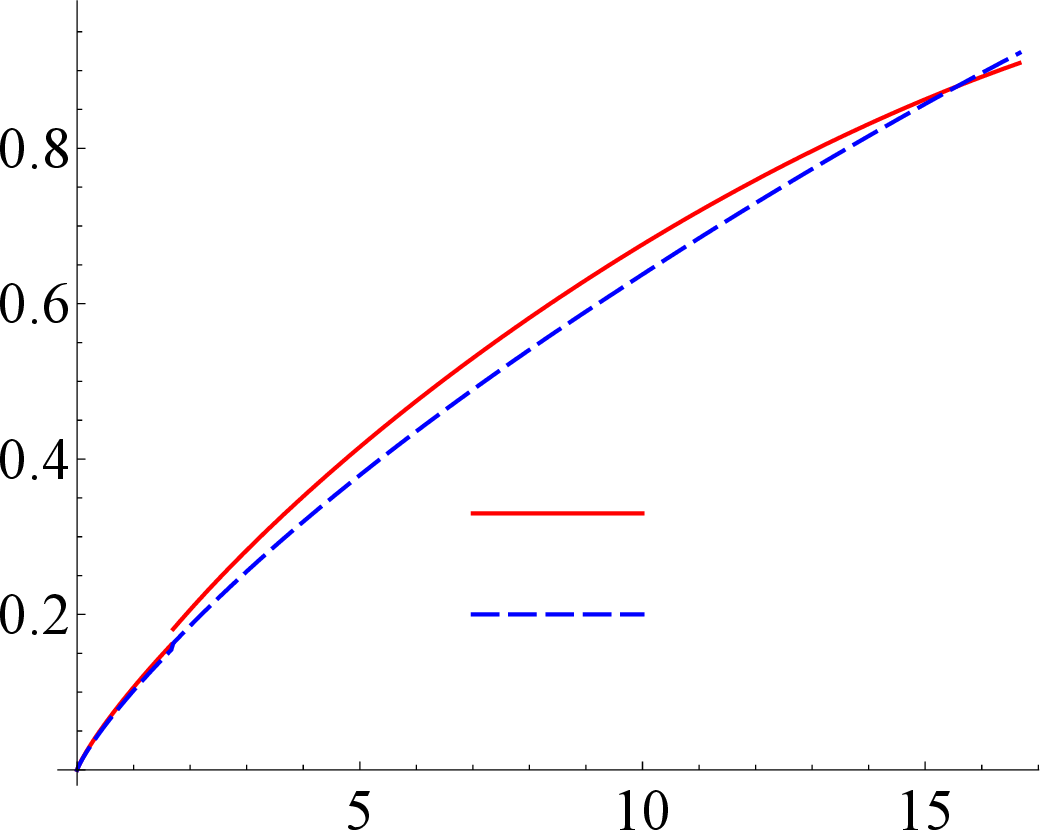}}}		
		\put(226,2){\resizebox{7.0 cm}{!}{\includegraphics{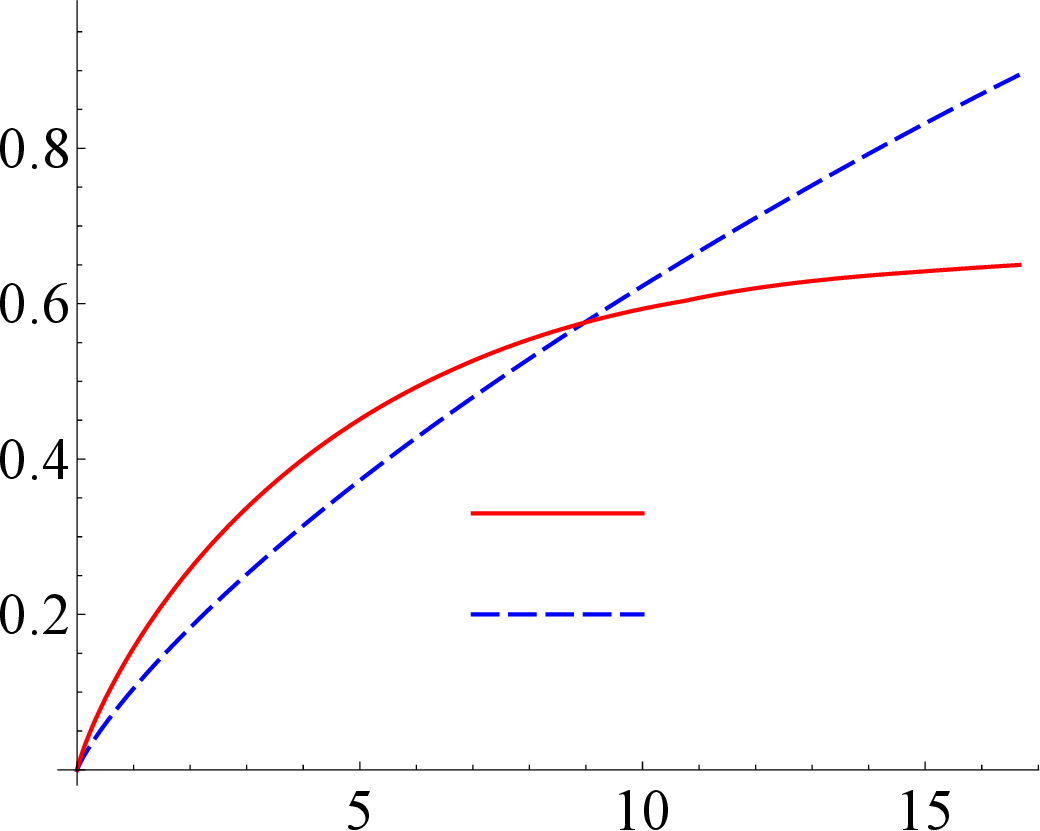}}}		
		\put(100,175){\textit{a}}
		\put(326,175){\textit{b}}
		\put(127,60){$\Gamma(\mathbf{B})-\Gamma(\mathbf{B}_1)$}
		\put(127,40){$\Gamma(\mathbf{B}_{\epsilon})$}
		\put(354,60){$\Gamma(\mathbf{B})-\Gamma(\mathbf{B}_4)$}
		\put(354,40){$\Gamma(\mathbf{B}_{\epsilon})$}
		\put(10,165){$\Gamma \ [\mbox{J/m}^2]$}
		\put(236,165){$\Gamma \ [\mbox{J/m}^2]$}
		\put(187,2){$\theta_{\epsilon} \ [\null^{\circ}]$}
		\put(414,2){$\theta_{\epsilon} \ [\null^{\circ}]$}
	\end{picture}
	\vskip 0.0cm
	\caption{Slopes of the BRK energy function for Ni  near cusps 
		at $\mathbf{B}_C = \mathbf{B}_1$ (\textit{a}) 
		and $\mathbf{B}_C = \mathbf{B}_4$  (\textit{b})
		compared to corresponding slopes of the cusp at $\mathbf{B}_0$.  
		The misorientation of $\mathbf{B}_{\epsilon}$ is
		$M_{\epsilon} = R \left(\mathbf{k}_{\epsilon}, \theta_{\epsilon} \right)$.
		The directions of $\mathbf{k}_{\epsilon}$ 
		are $[\overline{2}\, 0\, 1]$ in (\textit{a})
		and $[1\, 1\, 2]$  in (\textit{b}). 
		Values of remaining parameters are determined 
		by the geometric compatibility conditions.
	}
	\label{Fig_slopes_BRK_FA}
\end{figure}
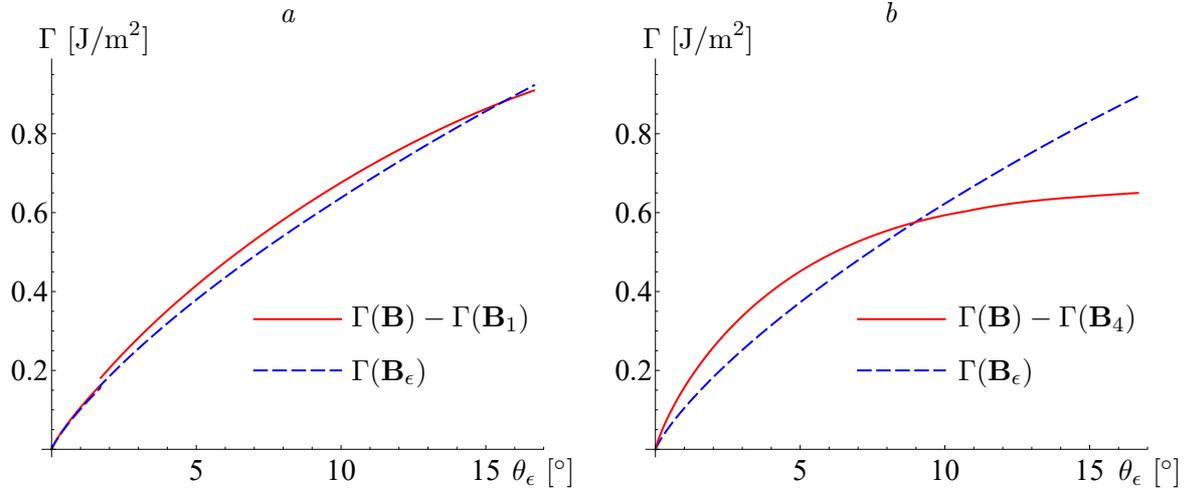

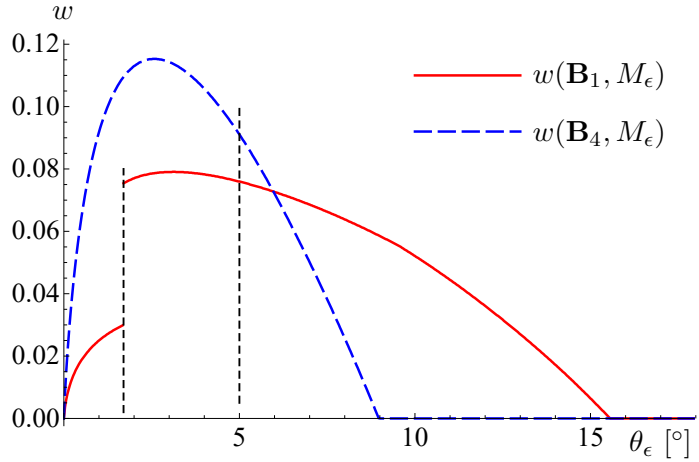
\begin{figure}
	\begin{picture}(300,170)(0,0)
		\put(80,2){\resizebox{9.0 cm}{!}{\includegraphics{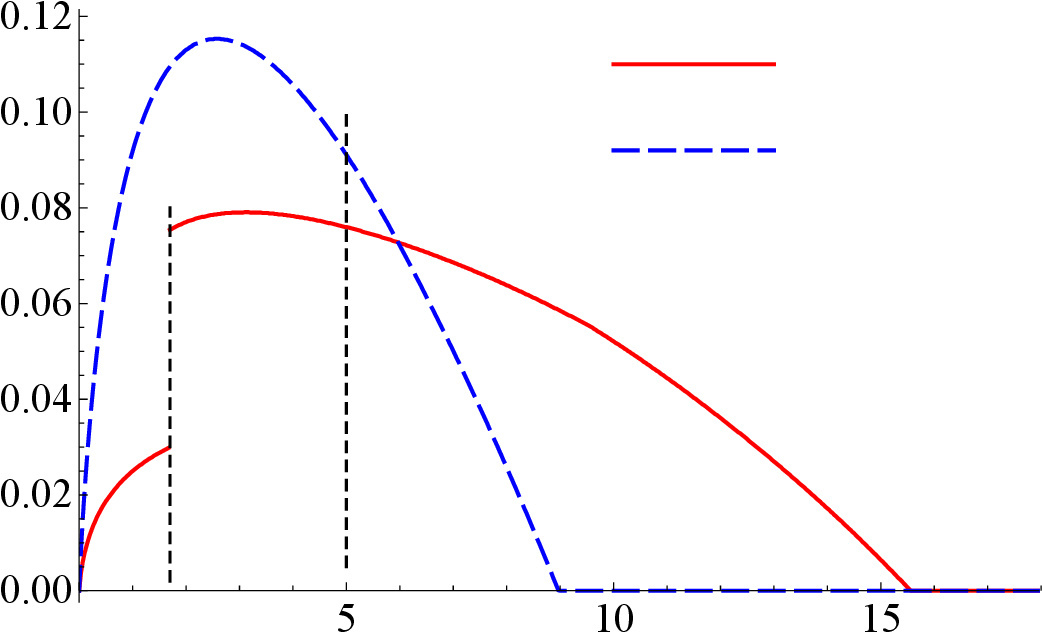}}}		
		\put(95,163){$w$}
		\put(275,139){$w(\mathbf{B}_1,M_{\epsilon})$}
		\put(275,117){$w(\mathbf{B}_4,M_{\epsilon})$}
		\put(311,0){$\theta_{\epsilon} \ [\null^{\circ}]$}
	\end{picture}
	\vskip 0.0cm
	\caption{Wettability parameter $w(\mathbf{B}_C,M_{\epsilon})$ versus 
		the misorientation angle $\theta_{\epsilon}$ 
		calculated using the BRK model for Ni. 
		The misorientation of $\mathbf{B}_{\epsilon}$ is
		$M_{\epsilon} = R \left(\mathbf{k}_{\epsilon}, \theta_{\epsilon} \right)$.
		The directions of $\mathbf{k}_{\epsilon}$ are 
		the same as in Fig.~\ref{Fig_slopes_BRK_FA}. 
		The discontinuity of $w(\mathbf{B}_1,M_{\epsilon})$ 
		at $\theta_{\epsilon} \approx 1.68^{\circ}$ 
		results from a discontinuity present in the BRK model.
	}
	\label{Fig_small_angle_BRK_FA}
\end{figure}

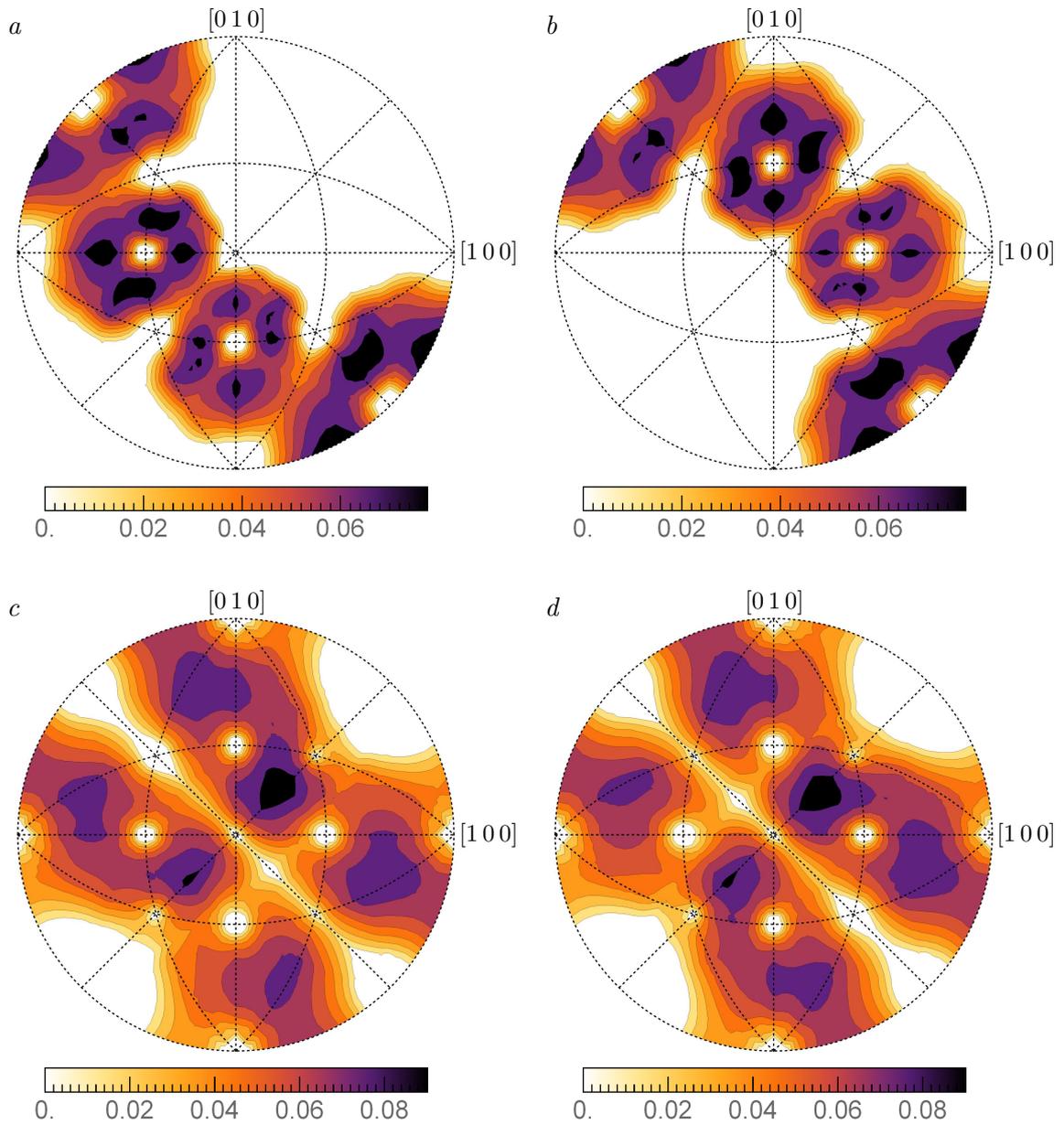
\begin{figure}
	\begin{picture}(300,460)(0,0)
		\put(20,240){\resizebox{6.6 cm}{!}{\includegraphics{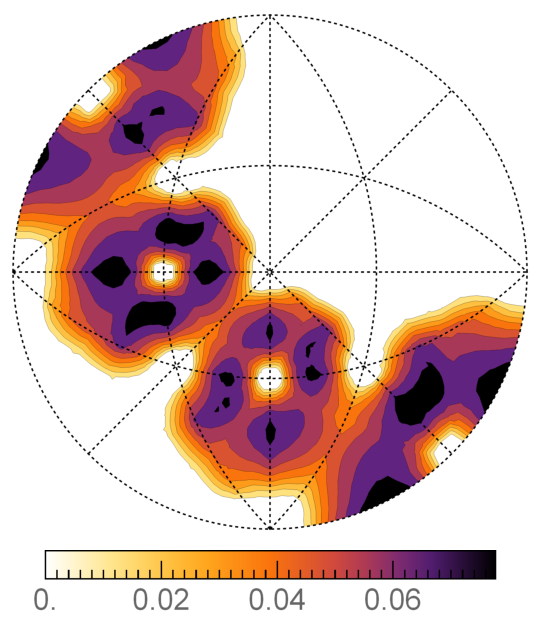}}}		
		\put(240,240){\resizebox{6.6 cm}{!}{\includegraphics{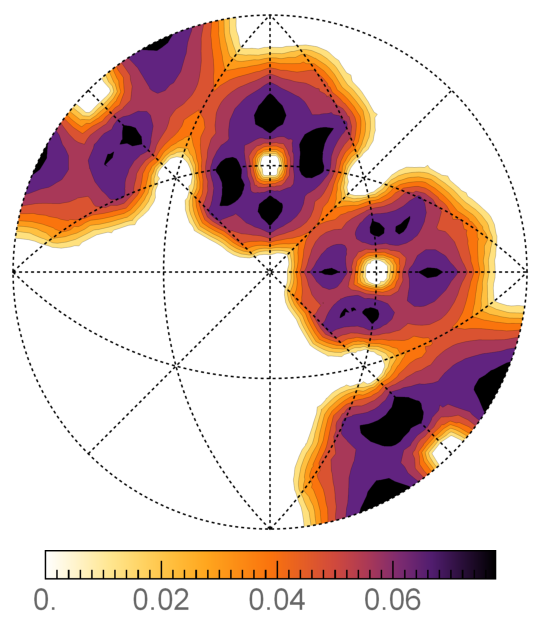}}}		
		\put(20,0){\resizebox{6.6 cm}{!}{\includegraphics{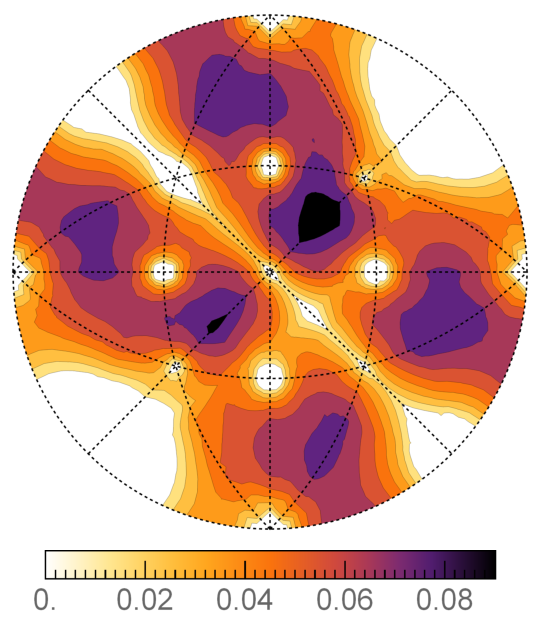}}}		
		\put(240,0){\resizebox{6.6 cm}{!}{\includegraphics{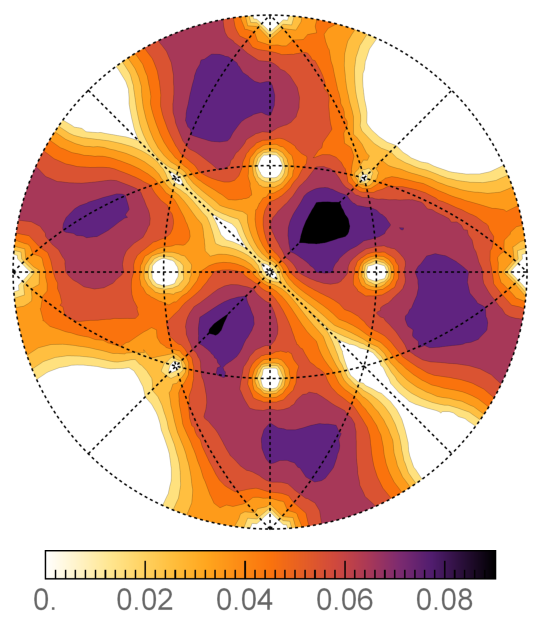}}}
		\put(20,450){\textit{a}}
		\put(240,450){\textit{b}}
		\put(205,357){\small $[ 1\, 0\, 0]$}
		\put(102,453){\small $[ 0\, 1\, 0]$}
		\put(425,357){\small $[ 1\, 0\, 0]$}
		\put(322,453){\small $[ 0\, 1\, 0]$}
		\put(20,210){\textit{c}}
		\put(240,210){\textit{d}}
		\put(205,117){\small $[ 1\, 0\, 0]$}
		\put(102,213){\small $[ 0\, 1\, 0]$}
		\put(425,117){\small $[ 1\, 0\, 0]$}
		\put(322,213){\small $[ 0\, 1\, 0]$}
	\end{picture}
	\vskip 0.0cm
	\caption{
		Wettability parameter $w(\mathbf{B}_C,M_{\epsilon})$
		versus the misorientation axis $\mathbf{k}_{\epsilon}$ of $M_{\epsilon}$
		based on the BRK model for Ni. 
		The angle of $M_{\epsilon}$ was set to $5^{\circ}$.
		Results for the cusps at the twin boundary 
		$\mathbf{B}_C = \mathbf{B}_1$ (\textit{a},\textit{b}) 
		and at 
		$\Sigma 11$ boundary  $\mathbf{B}_C = \mathbf{B}_4$ (\textit{c},\textit{d}).
		Figures (\textit{a},\textit{c}) and (\textit{b},\textit{d}) show stereographic projections of the upper and
		lower hemishperes from the poles 		
		$[0 \, 0 \, \overline{1}]$ and $[0 \, 0 \, 1]$,
		respectively.  
	}
	\label{Fig_small_angle_BRK}
\end{figure}

The BRK model for Ni also violates 
inequality~(\ref{eq:dewet_cond_small_angle}).
Of several cases tested, violations occurred 
near cusps at $\mathbf{B}_C=\mathbf{B}_1$, $\mathbf{B}_4$, $\mathbf{B}_5$, and $\mathbf{B}_6$.
With fixed $\mathbf{B}_C$, the misorientation of 
the low-angle boundaries $\mathbf{B}_{\epsilon}$
given by 
$M_{\epsilon} = R \left(\mathbf{k}_{\epsilon}, \theta_{\epsilon} \right)$, 
and values of the remaining parameters determined 
by the geometric compatibility conditions, 
$\Gamma(\mathbf{B})$ is larger than 
$\Gamma(\mathbf{B}_C)+ \Gamma(\mathbf{B}_{\epsilon})$
for some $\mathbf{k}_{\epsilon}$ and $\theta_{\epsilon}$.
Fig.~\ref{Fig_slopes_BRK_FA} shows example dependence of 
$\Gamma(\mathbf{B})-\Gamma(\mathbf{B}_C)$ and $\Gamma(\mathbf{B}_{\epsilon})$
on $\theta_{\epsilon}$ for 
$\mathbf{B}_C = \mathbf{B}_1$ and $\mathbf{B}_C = \mathbf{B}_4$ and 
selected vectors $\mathbf{k}_{\epsilon}$.
The figure compares the shapes of the cusps at $\mathbf{B}_C$ 
with the corresponding profiles of the cusp at $\mathbf{B}_0$.
The slopes in the vicinity of $\mathbf{B}_1$ 
and $\mathbf{B}_4$ are steeper than those near $\mathbf{B}_0$.
Susceptibility to splitting is more effectively characterized 
by the parameter $w$.
Fig~\ref{Fig_small_angle_BRK_FA} presents the dependence of 
$w$ on $\theta_{\epsilon}$, as computed from the data illustrated in 
Fig.~\ref{Fig_slopes_BRK_FA}.
The values of $w$ are positive not only for 
the selected directions of the vector $\mathbf{k}_{\epsilon}$, 
but also over a substantial portion of its domain.
The parameter $w$ as a function of $\mathbf{k}_{\epsilon}$ 
for $\theta_{\epsilon}=5^{\circ}$
and cusps at $\mathbf{B}_1$ and $\mathbf{B}_4$ 
is shown in Fig.~\ref{Fig_small_angle_BRK}. 

A question arises as to whether the violation of inequality (\ref{eq:dewetting_condition}) is a consequence of the fitting 
procedure used to define the BRK model, or whether it is inherent 
in the simulation data of Olmsted et al. \cite{olmsted2009survey}, 
upon which the model is based. 
It is the former: 
The dataset of Olmsted et al.  
contains 2664 triplets 
$\left( \mathbf{B}_a, \mathbf{B}_{b_1}, \mathbf{B}_{b_2} \right)$
satisfying the compatibility conditions (\ref{eq:compatibility_conditions}), 
but none of them violates 
inequality~(\ref{eq:dewetting_condition}).
The set includes a triplet equivalent to (\ref{eq_boundaries_S9}).
In this instance, the ratio 
$(\Gamma(\mathbf{B}_{b_1})+\Gamma(\mathbf{B}_{b_2}))/\Gamma(\mathbf{B}_a)$
for Ni is the lowest among all triplets 
satisfying~(\ref{eq:compatibility_conditions}), but 
with a value of 1.031, it still exceeds 1.\footnote{With the model 
described in \cite{Morawiec_2025a}
and Ni data from \cite{olmsted2009survey},
the boundaries 
(\ref{eq_boundaries_S9})
satisfy inequality (\ref{eq:dewetting_condition}). The 
model of \cite{Morawiec_2025a} exhibits unphysical 
behavior near $\mathbf{B}_0$, 
and there is no point in testing it against (\ref{eq:dewet_cond_small_angle}).}

In addition to Ni, the BRK model also applies to Cu, Al and Au.
In contrast to the cases of Ni, Cu and Au, the model for Al satisfies 
inequality (\ref{eq:dewetting_condition})
for the boundaries defined in (\ref{eq_boundaries_S9}).
A violation of (\ref{eq:dewet_cond_small_angle}) is 
observed near cusps at 
$\mathbf{B}_C=\mathbf{B}_4$, $\mathbf{B}_5$, and $\mathbf{B}_6$ for Al, 
and near $\mathbf{B}_1$ and $\mathbf{B}_4$ for Cu and Au.

From a physical viewpoint, 
the violation of (\ref{eq:dewetting_condition})
means that the BRK model allows for 
the decomposition of a boundary into two boundaries, and 
one might expect such events to occur in BRK-based grain growth simulations.
Within the continuum approximation, allowing boundary dissociation in an energy model is expected to influence microstructural evolution. 
Boundary types that undergo dissociation will tend to be underrepresented in the resulting microstructure.
It should be noted, however, that in practice splitting a boundary would require overcoming an energy barrier (associated with nucleating a new grain) that may exceed the energy gain obtained from the splitting.
Moreover, the boundary parallelism assumed above is a purely mathematical concept that is valid within the continuum approximation but has no counterpart in the actual process of grain boundary motion.
It is not surprising that no boundary splitting 
has been reported in BRK-based grain growth simulations 
\cite{hallberg2019modeling,nino2023influence,Naghibzadeh_2024,nino2024evolution}.

It is not implied here that an energy model must necessarily exclude boundary splitting. However, if a model is derived from data without boundary splitting, it may be reasonable to assume that the model inherits this property. For future work, new energy models should be tested to determine whether they permit boundary splitting.

Energy condition (\ref{eq:dewetting_condition}) also arises in connection with other phenomena, including general forms of boundary dissociation \cite{Goodhew_1978,Forwood_1984,Garg_1989}, twin formation \cite{Fullman_1951}, and boundary wetting or faceting \cite{Blendell_1999}. The analysis presented above addresses a simple case of boundary-plane geometry but, unlike more general treatments, this restriction allows for the incorporation of complete formal geometric constraints.

In summary, a structure can reduce its overall energy by replacing 
a single boundary with several boundaries whose total energy is 
lower than that of the original. 
Formal relationships describing the simple case of 
a planar boundary splitting into two parallel boundaries have been considered. 
These relationships can be used to test grain 
boundary energy models and determine whether they permit or prevent such splitting.
The tests help in understanding how the models perform in simulations of polycrystalline materials. 
The Bulatov–Reed–Kumar model has been shown to allow boundary splitting.

\clearpage

\bibliographystyle{unsrt}
\bibliography{GBEnergy.bib}

\begin{thebibliography}{10}

\bibitem{bulatov2014grain}
V.V. Bulatov, B.W. Reed, and M.~Kumar.
\newblock Grain boundary energy function for fcc metals.
\newblock {\em Acta Mater.}, 65:161--175, 2014.

\bibitem{sarochawikasit2021grain}
R.~Sarochawikasit, C.~Wang, P.~Kumam, H.~Beladi, T.~Okita, G.S. Rohrer, and
  S.~Ratanaphan.
\newblock Grain boundary energy function for $\alpha$ iron.
\newblock {\em Materialia}, 19:101186, 2021.

\bibitem{chirayutthanasak2024universal}
O.~Chirayutthanasak, R.~Sarochawikasit, S.~Khongpia, T.~Okita, S.~Dangtip, G.S.
  Rohrer, and S.~Ratanaphan.
\newblock Universal function for grain boundary energies in bcc metals.
\newblock {\em Scripta Mater.}, 240:115821, 2024.

\bibitem{Morawiec_2025a}
A.~Morawiec.
\newblock On modeling global grain boundary energy functions.
\newblock {\em Acta Mater.}, 286:120697, 2025.

\bibitem{Ratanaphan_2019}
S.~Ratanaphan, R.~Sarochawikasit, N.~Kumanuvong, S.~Hayakawa, H.~Beladi, G.S.
  Rohrer, and T.~Okita.
\newblock Atomistic simulations of grain boundary energies in austenitic steel.
\newblock {\em J. Mater. Sci.}, 54:5570--5583, 2019.

\bibitem{zhang2022molecular}
Y.~Zhang, E.D. Hansen, T.~Harbison, S.~Masengale, J.~French, and L.~Aagesen.
\newblock A~molecular dynamics survey of grain boundary energy in uranium
  dioxide and cerium dioxide.
\newblock {\em J. Am. Ceram. Soc.}, 105:4471--4486, 2022.

\bibitem{olmsted2009survey}
D.L. Olmsted, S.M. Foiles, and E.A. Holm.
\newblock Survey of computed grain boundary properties in face-centered cubic
  metals: {I}. {G}rain boundary energy.
\newblock {\em Acta Mater.}, 57:3694--3703, 2009.

\bibitem{Ratanaphan_2015}
S.~Ratanaphan, D.L. Olmsted, V.V. Bulatov, E.A. Holm, A.D. Rollett, and G.S.
  Rohrer.
\newblock Grain boundary energies in body-centered cubic metals.
\newblock {\em Acta Mater.}, 88:346--354, 2015.

\bibitem{Goux_1961}
C.~Goux.
\newblock Etude de la structure et des propri{\'e}t{\'e}s des joints de grains
  {\`a} l'aide de bicrystaux orient{\'e}s en aluminium pur.
\newblock {\em M{\'e}m. scient. Revue M{\'e}tall.}, 58:661--676, 1961.

\bibitem{Lange_1967}
F.F. Lange.
\newblock Mathematical characterization of a general bicrystal.
\newblock {\em Acta Metall.}, 15:311--318, 1967.

\bibitem{Fortes_1972}
M.A. Fortes.
\newblock Grain boundary parameters.
\newblock {\em Acta Cryst.}, A28:100--102, 1972.

\bibitem{saylor2003relative}
D.M. Saylor, A.~Morawiec, and G.S. Rohrer.
\newblock The relative free energies of grain boundaries in magnesia as a
  function of five macroscopic parameters.
\newblock {\em Acta Mater.}, 51:3675--3686, 2003.

\bibitem{beladi2014five}
H.~Beladi, N.T. Nuhfer, and G.S. Rohrer.
\newblock The five-parameter grain boundary character and energy distributions
  of a fully austenitic high-manganese steel using three dimensional data.
\newblock {\em Acta Mater.}, 70:281--289, 2014.

\bibitem{shen2019determining}
Y.F. Shen, X.~Zhong, H.~Liu, R.M. Suter, A.~Morawiec, and G.S. Rohrer.
\newblock Determining grain boundary energies from triple junction geometries
  without discretizing the five-parameter space.
\newblock {\em Acta Mater.}, 166:126--134, 2019.

\bibitem{Goodhew_1978}
P.J. Goodhew, T.Y. Tan, and R.W. Balluffi.
\newblock Low energy planes for tilt grain boundaries in gold.
\newblock {\em Acta Metall.}, 26:557--567, 1978.

\bibitem{Forwood_1984}
C.T. Forwood and L.M. Clarebrough.
\newblock Dissociation of asymmetric {$\Sigma 9$}, {$\Sigma 27a$} and {$\Sigma
  81d$} $\langle 110 \rangle$ tilt boundaries.
\newblock {\em Acta Metall.}, 32:757--771, 1984.

\bibitem{Sutton_1995}
A.P. Sutton and R.W. Balluffi.
\newblock {\em Interfaces in crystalline materials}.
\newblock Clarendon Press, Oxford, 1995, p. 402.

\bibitem{Morawiec_1998}
A.~Morawiec.
\newblock Symmetries of grain boundary distributions.
\newblock In B.~Adams, A.~Rollett, and H.~Weiland, editors, {\em Grain Growth
  in Polycrystalline Materials III}, pages 509--514, Warrendale, PA, US, 1998.
  TMS.

\bibitem{Morawiec_2009}
A.~Morawiec.
\newblock Models of uniformity for grain boundary distributions.
\newblock {\em J. Appl. Cryst.}, 42:783--792, 2009.

\bibitem{Bruckner_1962}
A.M. Bruckner and E.~Ostrow.
\newblock Some function classes related to the class of convex functions.
\newblock {\em Pac. J. Math.}, 14:1203--1215, 1962.

\bibitem{Morawiec_2019_AM}
A.~Morawiec.
\newblock On distances between grain interfaces in macroscopic parameter space.
\newblock {\em Acta Mater.}, 181:399--407, 2019.

\bibitem{hallberg2019modeling}
H.~Hallberg and V.V. Bulatov.
\newblock Modeling of grain growth under fully anisotropic grain boundary
  energy.
\newblock {\em Model. Simul. Mater. Sc.}, 27:045002, 2019.

\bibitem{nino2023influence}
J.D. Ni{\~n}o and O.K. Johnson.
\newblock Influence of grain boundary energy anisotropy on the evolution of
  grain boundary network structure during {3D} anisotropic grain growth.
\newblock {\em Comp. Mater. Sci.}, 217:111879, 2023.

\bibitem{Naghibzadeh_2024}
S.K. Naghibzadeh, Z.~Xu, D.~Kinderlehrer, R.~Suter, K.~Dayal, and G.S. Rohrer.
\newblock Impact of grain boundary energy anisotropy on grain growth.
\newblock {\em Phys. Rev. Mater.}, 8:093403, 2024.

\bibitem{nino2024evolution}
J.~Ni{\~n}o and O.K. Johnson.
\newblock Evolution of crystallographic texture and grain boundary network
  structure during anisotropic grain growth.
\newblock {\em Comp. Mater. Sci.}, 240:113023, 2024.

\bibitem{Garg_1989}
A.~Garg, W.A.T. Clark, and J.P. Hirth.
\newblock Dissociated and faceted large-angle coincident-site-lattice
  boundaries in silicon.
\newblock {\em Philos. Mag. A}, 59:479--499, 1989.

\bibitem{Fullman_1951}
R.L. Fullman and J.C. Fisher.
\newblock Formation of annealing twins during grain growth.
\newblock {\em J. Appl. Phys.}, 22:1350--1355, 1951.

\bibitem{Blendell_1999}
J.E. Blendell, W.~C. Carter, and C.A. Handwerker.
\newblock Faceting and wetting transitions of anisotropic interfaces and grain
  boundaries.
\newblock {\em J. Amer. Ceram. Soc.}, 82:1889--1900, 1999.

\end{thebibliography}

\end{document}